\documentclass[preprint,sort]{elsarticle}

\usepackage{subfloat}
\usepackage{xcolor}
\usepackage{multirow}
\usepackage{amssymb, amsmath}
\usepackage[parfill]{parskip}
\usepackage[export]{adjustbox}
\usepackage{graphicx, subfigure}
\usepackage{hyperref}
 \hypersetup{
 colorlinks = true,
 allbordercolors = {white},
 allcolors = {blue},
 }
\biboptions{sort&compress}

\def\url#1{}

\begin{document}
 \begin{frontmatter}
 \title{Ultrasound Triggering of Rayleigh-Taylor Instability: Solution of Compressible Navier-Stokes Equation by a Non-Overlapping Parallel Compact Scheme}
 \author{Prasannabalaji Sundaram$^a*$, Aditi Sengupta$^b$, Tapan K. Sengupta$^{b}$}
 \address{$^a$ High Performance Computing Laboratory,
 Department of Aerospace Engineering, \\
 I.I.T. Kanpur, Kanpur 208016, India \\
 $^b$ Department of Mechanical Engineering \\
 I.I.T. (ISM) Dhanbad, Jharkhand 826004, India \\
 *Email: prasanna@iitk.ac.in}

\begin{abstract}
Rayleigh-Taylor instability (RTI) occurs at the interface of two media when the heavier fluid is accelerated into the lighter fluid and is a prototypical hydrodynamic event present in many physical events. In high energy physics, this manifests itself across a wide range of length scales from nuclear confinement fusion at micron-scale to supernova explosion at terra scales. RTI can also be viewed as a baroclinic instability prevalent in engineering, geophysics, and astrophysics, a pedagogic description of which is given in Sengupta {\it et al., Comput. Fluids,} {\bf 225}, 104995 (2021) with respect to the experimental results of Read, {\it Physica D}, {\bf 12} 45-58 (1984). Here, a recently proposed non-overlapping parallel algorithm is used to solve this three-dimensional canonical problem, having the unique property of not distinguishing between sequential and parallel computing, using 4.19 billion points and a refined time step of $7.69 \times 10^{-8} sec$. The problem achieves the required density gradient by considering two volumes of air at different temperatures (with a temperature difference of 200K) separated by a non-conducting, impermeable partition at the onset of the experiment, which is removed impulsively at $t=0$. The resulting buoyancy force at the interface acting from top to bottom is the seed of the baroclinic instability. Present high precision computation enables one to capture the ensuing RTI triggered by ultrasonic waves created at the interface.

\end{abstract}


\end{frontmatter}

\section{Introduction}
The Rayleigh-Taylor instability (RTI) \cite{B125_4, B125_5} is present in different branches of physics, from hydrodynamic applications \cite{B125_22, B125_30,B125_25, Robey_etal_Angulo} to high energy physics of astrophysics \cite{Cabot_Cook_Nature,Remington_etal_Angulo} and nuclear confinement fusion \cite{B125_3,Nagel_etal_Angulo}. Similarly, the baroclinic instability can arise during onset of RTI due to misalignment of density and pressure gradients, a major source of torque that contributes to the instability \cite{TT_CUP, B125}. In the present research, the compressible flow formulation of the Navier-Stokes equation (NSE) in \cite{B125} has been used. Such a formulation obviates the need of making the Boussinesq approximation, as needed for incompressible flow formulations. This has been critiqued by Mikaelian \cite{Mikaelian14} highlighting the need for using compressible formulation. The baroclinic torque term is directly evident from the inviscid vorticity transport equation \cite{TT_CUP} as,

\begin{equation}
\frac{D\vec{\omega}}{Dt} \equiv \frac{\partial \vec{\omega}}{\partial t} + \biggl(\vec{V} \cdot \nabla\biggr)\vec{\omega} = \biggl(\vec{\omega} \cdot \nabla\biggr)\vec{V} - \vec{\omega} \biggl(\nabla \cdot \vec{V}\biggr) + \frac{1}{\rho^2} \nabla\rho \times \nabla p
\label{eqn12_1}
\end{equation}

\noindent where $\vec{V}$ and $\vec{\omega}$ are the velocity and vorticity vectors, and the last term on the right-hand side ($\frac{1}{\rho^2} \nabla\rho \times \nabla p$) is the baroclinic contribution to vorticity generation by the misaligned pressure and density gradients. For baroclinic flows, the pressure is not dependent on density alone, making this contribution non-zero. The term given by, $\vec{\omega} \biggl(\nabla \cdot \vec{V}\biggr)$, is due to compressibility, and RTI should justifiably depend upon bulk viscosity as noted from the constitutive relation between stress and rates of strain tensors. Equation \eqref{eqn12_1} demonstrates the need for predicting RTI using the compressible flow formulation along with the bulk viscosity contribution, as has been reported in \cite{B125_21, B125}, and the same formulation is also used here. Thus, without using the Stokes' hypothesis, the linear regression model for the bulk viscosity based on acoustic dispersion and attenuation experimental data of Ash {\it et al.} \cite{B125_46} is utilized here also.

\begin{figure*}
 \centering
 \includegraphics[width=0.75\textwidth]{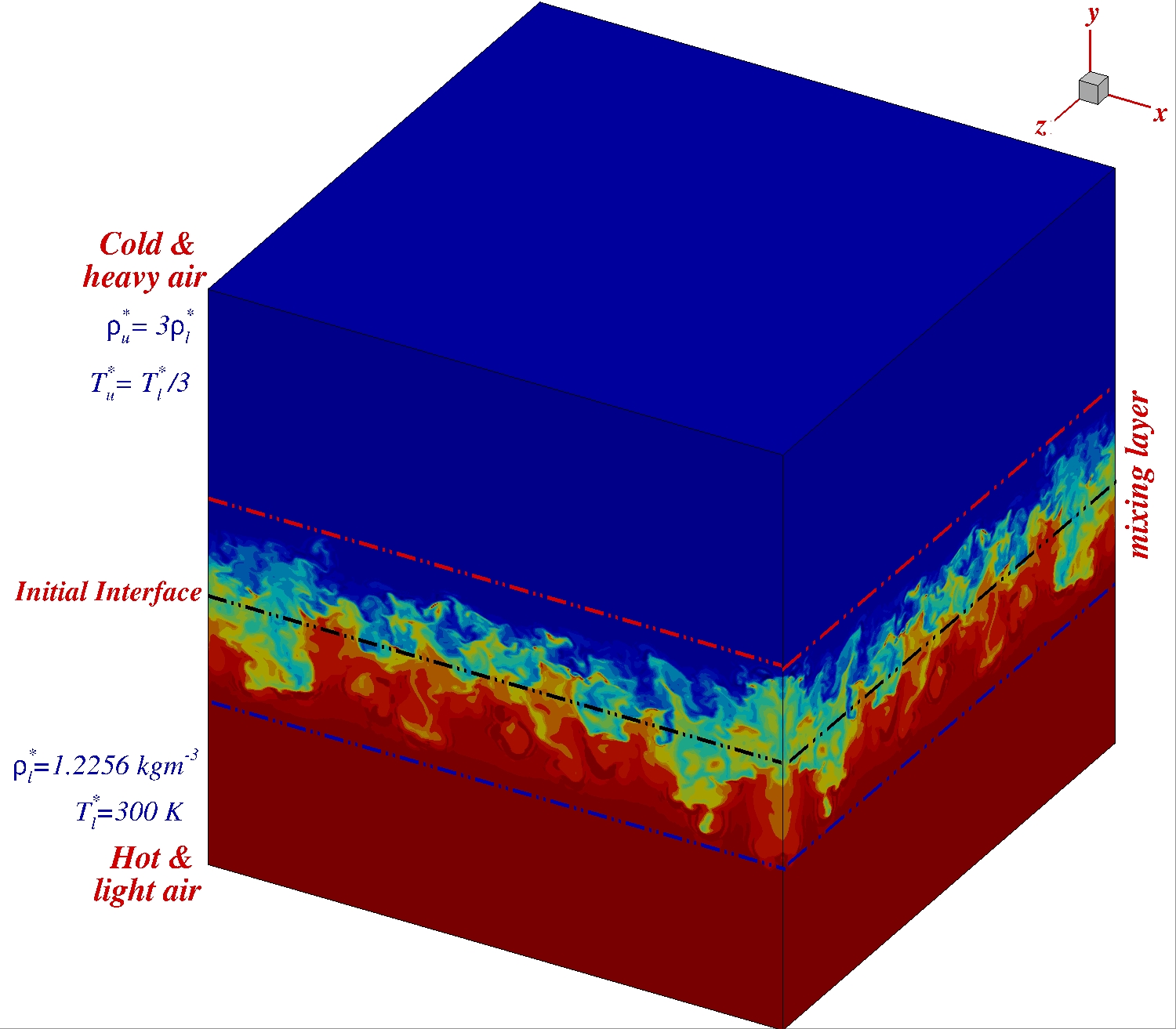}
 \caption{Schematic of the RTI problem solved inside a box having square cross section with initial conditions indicating the heavy fluid resting atop the lighter fluid.}
 \label{Fig:Schematic}
\end{figure*}

Buoyancy-induced acceleration due to gravity has been considered in \cite{B125_10, B125_11, B125_12, B125_13} for a tank with heavy fluid resting atop a lighter fluid, initially separated by a horizontal partition, removal of which causes RTI. A schematic of the physical domain along with initial conditions on either side of the interface is shown in Fig. \ref{Fig:Schematic}. Such experiments allow detailed observation of the evolving mixing layer after the partition is removed at $t =0$, quite unlike those experiments where RTI is triggered by the acceleration of the chamber. In the present simulations, a similar configuration for RTI is studied, with large temperature difference (200K) between the top and bottom fluids causing the Atwood number ($=\frac{\rho_u - \rho_l}{\rho_u + \rho_l}$) to be $A_t = 0.5$. Air is the working fluid that is treated as a perfect gas. In the present case, the box is considered to be thermally insulated, which enables it to be viewed as an {\it universe}, without any mass, momentum, and energy transfer with the outside. The ensuing RTI upon removal of the partition enables one to view the process as one of non-equilibrium thermodynamics, as reported for a two-dimensional (2D) flow in \cite{B125_20, B125_21, B125_27} and a three-dimensional (3D) flow with $A_t = 0.2$, due to initial temperature differential of 149K. In Read's experiments \cite{B125_7}, two geometrical setups were considered with the purpose of providing results for 2D and 3D flows. The present research aims to consider only the 3D RTI setup with $A_t = 0.5$, while in \cite{B125}, corresponding 2D and 3D setups were computed by solving the 3D geometry with cross-sections of $(0.15m \times 0.025m)$ and $(0.15m \times 0.15m)$, respectively.

As implied in linear instability theory \cite{B125_5,B125_6, IFTT}, RTI is caused at the interface of two fluids with dissimilar density due to three main parameters:
(i) the imposed acceleration that causes the dense fluid (with density, $\rho_u$) to penetrate into the lighter fluid ($\rho_l$) as a spatio-temporal event;
(ii) a non-dimensional density ratio given by the Atwood number ($A_t$) and
(iii) a length scale given by a real wavenumber ($\alpha$). In fact, the dispersion relation is in general complex, but in many experimental \cite{Nagel_etal_Angulo}, and numerical investigations \cite{B125_14, Clark_etal_Angulo, B125_22} the length scale is considered real, which makes the linear disturbance growth temporal. In general, the transition process will be due to a spatio-temporal mechanism.

In practice, there is no need to impose any length scale as there are many 2D and 3D direct numerical simulations (DNS) where the initial interface is purely taken as planar \cite{B125_20, B125_21, B125_27, B125_29, B125} to replicate the experiments reported in \cite{B125_7, B125_8, B125_11} where the fluid interface is initially plane and the onset of RTI has been shown to originate from the edges and corners of the initial interface \cite{B125}. In contrast, many other simulations required the onset of the RTI to be at lower wavenumbers due to numerical reasons. The issue of numerical and experimental space-time resolution has been explained in \cite{Angulo_etal}, highlighting the improved experimental capabilities in recent times while discussing the parameter ranges of the Liepmann-Taylor scale also in the context of the design of capsules for inertial confinement fusion \cite{B125_2}. The fuel mixture is bombarded by a laser for the initiation process of fusion. RTI occurs at two instances: once during the initial implosion of the target and when the high-temperature gas mixture is decelerated in encountering the colder outer layer. The turbulent mixing during the RTI brings a colder outer layer of fuel to the core and, in the process, can suppress the ignition at the core \cite{B125_3}. The turbulent mixing following RTI is significantly influenced by the created vorticity field, not only during nuclear fusion but also during a volcanic eruption and in other geophysical events. Thus, the role of length scales in RTI is important, and it has been experimentally investigated in \cite{B125_14} by studying the distinction between experiments with and without imposed length scales. The authors in \cite{B125} have compared the length scale of the spikes (vortical structures originating from the heavier to lighter fluid) in the experiments reported in \cite{B125_7, B125_8, B125_10, B125_11, B125_14}.

The above discussion indicates that the onset of RTI and associated length scales are still not unambiguously resolved. We have already noted that high accuracy compact schemes used for solving compressible NSE without the Stokes' hypothesis enable one to track the onset of RTI to be from the edges and corners of the interface \cite{B125_20, B125_21, B125_27} for 2D RTI problem. The authors in \cite{Angulo_etal} used a 2D multi-physics simulation package, HYDRA \cite{Langer_etal, Marinak_etal} with a spatial resolution of the order of $3 \mu m$ and time steps of the order of $100 ps$. It was noted that for the nuclear fusion experiments using radiography, the ideal space-time resolution should be of the order of $2 \mu m$ and $0 ps$ to capture the evolving spikes! In contrast, the authors in \cite{B125} have used various grids with a representative one using $(320 \times 640 \times 320)$ in the computational box of dimension of $(150mm \times 150mm \times 150mm)$ i.e. about 65.536 million points with 160 cores. The results obtained matched the mixing layer and growth rate data of Read \cite{B125_7}. In the present research, we report results obtained using $(1280 \times 2560 \times 1280)$ i.e. 4.19 billion points for the same 3D setup of \cite{B125_7} using 19,200 cores. Thus, the spatial resolution has improved significantly, and the time step has been refined to $7.69 \times 10^{-8} sec$ (with a non-dimensional time step of $6.25 \times 10^{-7}$, as compared to a time step of $10^{-6}$ in \cite{B125}. More importantly, a novel parallel algorithm (non-overlapping high accuracy parallel (NOHAP) compact scheme) proposed in \cite{B124} is used. The parallelization scheme ensures the removal of any error originating at the sub-domain boundaries due to boundary closure schemes mandatory for compact schemes. Furthermore, the Schwartz domain decomposition method in \cite{B125}, required overlap points at the subdomain boundaries following the parallelization method introduced in \cite{B39}, to avoid errors originating due to parallelization. The NOHAP scheme used in the present research has been used earlier for different problems of fluid mechanics, as in \cite{SumanSivaTekrBhau19, B123, SenguptaRoyChakSeng21}.






The objective of the present research is to investigate the onset of RTI by the parallel algorithm by reducing computational errors down to the level of equivalent sequential computing of the same problem. The manuscript is formatted in the following manner. In the next section, we describe the formulation and various parameters used to solve the RTI problem. In section 3, a brief description of the parallel computing strategy and the associated numerical method is presented for the RTI, showing the linear scaling property of the parallelization. In section 4, the onset of RTI is shown with the appearance of acoustic waves and their propagation. The properties of such ultrasonic waves are explored. This is followed by section 5, where numerical solutions are shown to establish the relationship of the acoustic trigger at the onset with the vortical development that leads to the evolution of the mixing layer. The paper closes with a summary and conclusions in section 6.
%
%
\section{Formulation of compressible Navier-Stokes equation for RTI}
\vspace{-2mm}
Here, computations are performed in a 3D box as shown in Fig. \ref{Fig:Schematic} of each side having a length ($L$) of 0.15m. This domain is spatially discretized with uniform spacing having 1280 points in the horizontal ($x^*,z^*$)-plane. In the plane perpendicular to the interface, twice as many points are considered. This consists of air at two different constant temperatures separated initially ($t = 0$) by an insulated partition, marked in the figure. The hot and lower compartment temperature is given as $T^*_l$ or $T_s$ equal to 300K, while the upper compartment air has the temperature given by $T^*_u = 100K$. For the perfect gas as working fluid in both the chambers, provides the Atwood number of $A_t = 0.5$. The walls of the box are adiabatically insulated, making the box a thermodynamically isolated system. At $t =0$ the partition is removed impulsively, the ensuing RTI is triggered by multiple causes, and the purpose of the present research is to identify the role of pressure pulses created right from the onset. According to Chandrasekhar \cite{B125_6}, generation of baroclinic torque at the junction of the walls with the interface is the prime mover of RTI. This has also been supported by experimental visualization in \cite{B125_1, B125_7}. The authors in \cite{B125_21} reported 2D RTI with the presence of spatio-temporal pressure fronts right at the onset. In the present investigation, the role of the initial acoustic signals is explored in identifying the prime cause in creating RTI. As the instability is due to the unstable arrangement of heavy fluid resting over the light fluid, one can define the velocity and the time scale to be given by,
$$U_s = \sqrt{gL} ~~~\text{ and }~~~ t_s = \sqrt{L/g},$$
\noindent which becomes 12.131 m/s and 0.1237 s, respectively. With the density of air at $T_s = $ 300K taken as $\rho_s = 1.2256 kg/m^3$, the reference Reynolds number and Mach number are $Re =12,080.6$ and $M = 0.0034939$. The stratification of density and temperature in the flow field at $t=0$ in experiments, such as in \cite{B125_7,
B125_10, B125_13, B125_14}, creates the initial perturbation, immediately after the removal of the partition, exciting all possible spatial and temporal scales by the destabilizing potential energy \cite{B125_4}. The 3D computations in \cite{B125} and here follow this aspect of unforced experiments wherein the RTI onset is due to disturbances applied on the equilibrium flow, which in this case is given by the quiescent condition. To achieve this, an appropriate formulation without unphysical assumptions is necessary. Thus, the aim here is to investigate an isolated system using well-calibrated numerical methods with extreme accuracy.

The unsteady 3D compressible NSE has been solved, as given by the set of partial differential equations expressed in the divergence form \cite{B125_48, B125_28} as,
\begin{equation}
\frac{\partial \hat Q}{\partial t^*} + \frac{\partial \hat E_c}{\partial x^*} + \frac{\partial \hat F_c}{\partial y^*}+ \frac{\partial \hat G_c}{\partial z^*} = \frac{\partial \hat E_v}{\partial x^*} + \frac{\partial \hat F_v}{\partial y^*} + \frac{\partial \hat G_v}{\partial z^*} + \hat S \label{ge1}
\end{equation}
\noindent with the conserved variables given as
\begin{equation}
 \hat Q = \left[\rho^*;~~~ \rho^* u^*;~~~ \rho^* v^*;~~~ \rho^* w^*;~~~ \rho^* e^*_t \right]^T \label{ge2}
\end{equation}

The convective fluxes are $\hat{E}$, $\hat{F}$, $\hat{G}$, and given as
\begin{eqnarray}
 \hat E_c &=& \left[ \rho^* u^*;~~~ \rho^* u^{*2} + p^*;~~ \rho^* u^* v^*;~~~~~ \rho^* u^* w^*;~~~~~~~ \left(\rho^* e^*_t + p^*\right)u^* \right]^T \label{ge3} \\
 \hat F_c &=& \left[ \rho^* v^*;~~~ \rho^* u^* v^*;~~~~~~~ \rho^* v^{*2} + p;~ \rho^* v^* w^*;~~~~~~~ \left(\rho^* e^*_t + p^*\right)v^* \right]^T \label{ge4} \\
 \hat G_c &=& \left[ \rho^* w^*;~~~ \rho^* u^* w^*;~~~~~~ \rho^* v^*w^*;~~~ \rho^* w^{*2} + p^*; ~~ \left(\rho^* e^*_t + p^*\right)w^* \right]^T \label{ge5}
\end{eqnarray}

Similarly, the viscous flux vectors are denoted as $\hat{E}_v$, $\hat{F}_v$, $\hat{G}_v$, which are given as
\begin{eqnarray}
 \hat E_v &=& \left[ 0;~~~ \tau^*_{xx};~~~ \tau^*_{xy};~~~ \tau^*_{xz};~~~ u^*\tau^*_{xx} + v^*\tau^*_{xy}+ w^*\tau^*_{xz}-q^*_x \right]^T \label{ge6} \\
 \hat F_v &=& \left[ 0;~~~ \tau^*_{yx};~~~ \tau^*_{yy};~~~ \tau^*_{yz};~~~ u^*\tau^*_{yx} + v^*\tau^*_{yy}+ w^*\tau^*_{yz}-q^*_y \right]^T \label{ge7} \\
 \hat G_v &=& \left[ 0;~~~ \tau^*_{zx};~~~ \tau^*_{zy};~~~ \tau^*_{zz};~~~ u^*\tau^*_{zx} + v^*\tau^*_{zy}+ w^*\tau^*_{zz}-q^*_z \right]^T \label{ge8}
\end{eqnarray}

\noindent The source term $\hat S$ is given by,
\begin{equation}
 \hat S = \left[0;~~~ 0;~~~ -\rho^* g;~~~ 0;~~~ -\rho^* v^* g \right]^T, ~~~ \text{with } g=9.81ms^{-2}. \label{ge8a}
\end{equation}

In the above, the variables $\rho^*$, $p^*$, $u^*$, $v^*$, $w^*$, $T^*$ and $e_t^*$ are the dimensional density, pressure, Cartesian components of fluid velocity, absolute temperature and specific internal energy, respectively. The specific heat ratio ($\gamma$) is equal to 1.4 for air as assembly of diatomic molecules. The stress tensor are $\tau^*_{ij}$, for $i, j =1; to; 3$, which are related to the strain rates given by the gradients of velocity as \cite{B125},
\begin{equation}
 \tau^*_{xy} = \tau^*_{yx} = \mu^* \left[ \frac{\partial v^*}{\partial x^*} + \frac{\partial u^*}{\partial y^*}\right];~ \tau^*_{xx} = \left[2\mu^* \frac{\partial u^*}{\partial x^*} + \lambda^* \left[ \frac{\partial u^*}{\partial x^*} + \frac{\partial v^*}{\partial y^*} + \frac{\partial w^*}{\partial z^*} \right] \right] \label{ge8b}
\end{equation}
\begin{equation}
 \tau^*_{yz} = \tau^*_{zy} = \mu^* \left[ \frac{\partial v^*}{\partial z^*} + \frac{\partial w^*}{\partial y^*}\right];~ \tau^*_{yy} = \left[2\mu^* \frac{\partial v^*}{\partial y^*} + \lambda^* \left[ \frac{\partial u^*}{\partial x^*} + \frac{\partial v^*}{\partial y^*} + \frac{\partial w^*}{\partial z^*} \right] \right] \label{ge9}
\end{equation}
\begin{equation}
 \tau^*_{xz} = \tau^*_{zx} = \mu^* \left[ \frac{\partial w^*}{\partial x^*} + \frac{\partial u^*}{\partial z^*}\right];~ \tau^*_{zz} = \left[2\mu^* \frac{\partial w^*}{\partial z^*} + \lambda^* \left[ \frac{\partial u^*}{\partial x^*} + \frac{\partial v^*}{\partial y^*} + \frac{\partial w^*}{\partial z^*} \right] \right] \label{ge10}
\end{equation}

The specific heat capacity ($C_v$) and thermal conductivity ($\kappa$) are used as constants here. The heat conduction terms $q^*_x$, $q^*_y$ and $q^*_z$ are given as
\begin{equation}
q^*_x = - \kappa \frac{\partial T^*}{\partial x^*}, \; q^*_y = - \kappa \frac{\partial T^*}{\partial y^*}, \; q^*_z = - \kappa \frac{\partial T^*}{\partial z^*}
\label{ge11}
\end{equation}

\noindent Stokes' hypothesis relates molecular viscosity, $\mu^*$ with the second coefficient of viscosity, $\lambda^*$ via the relation: $\lambda^* = -2\mu^*/3$. In the present research, the bulk viscosity is incorporated via the experimental data \cite{B125_46}, as described below. Sutherland's law is used for the viscosity as a function of temperature. Additionally, the ideal gas relation provides the constitutive equation as,
\begin{equation}
p^*=\rho^* R^*T^*,
\label{ge12}
\end{equation}
\noindent which in turn defines specific energy, $e_t^*$ as
\begin{equation}
e_t^* =\frac{p^*}{\rho^*(\gamma-1)} + \frac{u^{*2} + v^{*2} + w^{*2}}{2}
\label{ge13}
\end{equation}

We follow the procedure used in \cite{B125} to nondimensionalize the governing equations, with the variables related to appropriate scales as,
\begin{equation}
x = \frac{x^*}{L}, \; y = \frac{y^*}{L}, \; z = \frac{z^*}{L}, \; u = \frac{u^*}{U_s}, \; v = \frac{v^*}{U_s}, \; w = \frac{w^*}{U_s}, \; t = \frac{t^*U_s}{L},
\label{ge14}
\end{equation}
\begin{equation}
\rho = \frac{\rho^*}{\rho_s}, \; T = \frac{T^*}{T_s}, \; p = \frac{p^*} {p_s}, \; e_t = \frac{e^*_t}{U^2_s}, \; \mu = \frac{\mu^*}{\mu_s}
\label{ge15}
\end{equation}

\noindent where, $p_s = \rho_s R^* T_s$ is the characteristic pressure, and obtained with the reference temperature and density for hot air. Here, we have used $Pr = 0.712$, $\Delta T$ = 200 K, to match the experimental conditions in \cite{B125_7}. The nondimensionalized governing equations are obtained as
\cite{B125},
\begin{equation}
\frac{\partial \hat{Q}}{\partial t}+\frac{\partial \hat{E}}{\partial x}+\frac{\partial \hat{F}}{\partial y}+\frac{\partial \hat{G}}{\partial z}=\frac{\partial \hat{E_v}}{\partial x}+\frac{\partial \hat{F_v}}{\partial y}+\frac{\partial \hat{G_v}}{\partial z} + \hat S
\label{ge17}
\end{equation}

\noindent The off-diagonal terms of stress tensors are given by
\begin{eqnarray}
\tau_{xy} = \tau_{yx} &=& \frac{\mu}{Re} \left[\frac{\partial u}{\partial y} + \frac{\partial v}{\partial x} \right],
\tau_{xz} = \tau_{zx} = ~\frac{\mu}{Re} \left[\frac{\partial u}{\partial z} + \frac{\partial w}{\partial x} \right], \nonumber \\
\tau_{yz} = \tau_{zy} &=& \frac{\mu}{Re} \left[\frac{\partial w}{\partial y} + \frac{\partial v}{\partial z} \right] \label{ge19}
\end{eqnarray}

\noindent The heat conduction terms are given by
\begin{equation}
\left[q_x;~q_y;~q_z\right] = \frac{-\mu}{(\gamma-1) Pr Re M^2} \left[ \frac{\partial T}{\partial x}; \frac{\partial T}{\partial y}; \frac{\partial T}{\partial z} \right]
\label{ge20}
\end{equation}

\noindent The equation of state can be written as $p = \rho R T $, where $R = R^* T_s / U_s^2$.

There are in-depth discussions on the concept of the bulk viscosity and on the validity of Stokes' hypothesis in \cite{B125}. The authors have related the second coefficient of viscosity ($\lambda$) with the dynamic viscosity ($\mu$), by the expression $\lambda^* + \frac{2}{3} \mu^* = \mu_b$, where $\mu_b$ is the bulk viscosity of the fluid. Stokes \cite{B125_38} noted that in an analysis, with and without the bulk viscosity, if it produces different results, then it can be due to non-negligible values of divergence of velocity $\nabla \cdot \vec{V}$ contributing, along with $\mu_b$ being zero. For compressible flows, $\nabla \cdot \vec{V}$ is a significant non-negligible quantity. Past simulations of RTI \cite{B125, B125_21, B125_29} have shown the necessity to include the bulk viscosity to capture the early time behavior of RTI. Here, such an approach based on the acoustic dispersion and attenuation measurements in \cite{B125_46} for $\mu_b$ is adopted. Ash {\it et al.}'s experimental measurements provide values of $\mu_b$ for air as a function of temperature. Using a linear regression analysis of the experimental data, a relationship is drawn between $\mu_b$ and $T^*$ (in K) as \cite{B125},
\begin{equation}
 \mu_b = \frac{1}{10^4}\left[ 3.381T^* -7.383 \right]
 \label{reg}
\end{equation}

Thus, the non-dimensional normal stress components are obtained as \cite{B125},
\begin{equation}
\tau_{xx} = \frac{1}{Re} \biggl [ \biggl ( \frac{4}{3} \mu + \frac{\mu_b}{\mu_s} \biggr ) \frac{\partial u}{\partial x} + \biggl (-\frac{2}{3} \mu+ \frac{\mu_b}{\mu_s} \biggr ) \frac{\partial v}{\partial y} + \biggl (-\frac{2}{3} \mu + \frac{\mu_b}{\mu_s} \biggr ) \frac{\partial w}{\partial z} \biggr ]
\label{bv1}
\end{equation}
\begin{equation}
\tau_{yy} = \frac{1}{Re} \biggl [ \biggl ( \frac{4}{3} \mu + \frac{\mu_b}{\mu_s} \biggr ) \frac{\partial v}{\partial y} + \biggl (-\frac{2}{3} \mu + \frac{\mu_b}{\mu_s} \biggr ) \frac{\partial u}{\partial x} + \biggl (-\frac{2}{3} \mu + \frac{\mu_b}{\mu_s} \biggr ) \frac{\partial w}{\partial z} \biggr ]
\label{bv2}
\end{equation}
\begin{equation}
\tau_{zz} = \frac{1}{Re} \biggl [ \biggl ( \frac{4}{3} \mu + \frac{\mu_b}{\mu_s} \biggr ) \frac{\partial w}{\partial z} + \biggl (-\frac{2}{3} \mu + \frac{\mu_b}{\mu_s} \biggr ) \frac{\partial u}{\partial x} + \biggl (-\frac{2}{3} \mu + \frac{\mu_b}{\mu_s} \biggr ) \frac{\partial v}{\partial y} \biggr ]
\label{bv3}
\end{equation}
%

\section{An Accuracy Preserving Sub-domain Boundary Closure for Compact Scheme}

\begin{figure*}[h!]
 \centering
 \includegraphics[width=0.8\textwidth]{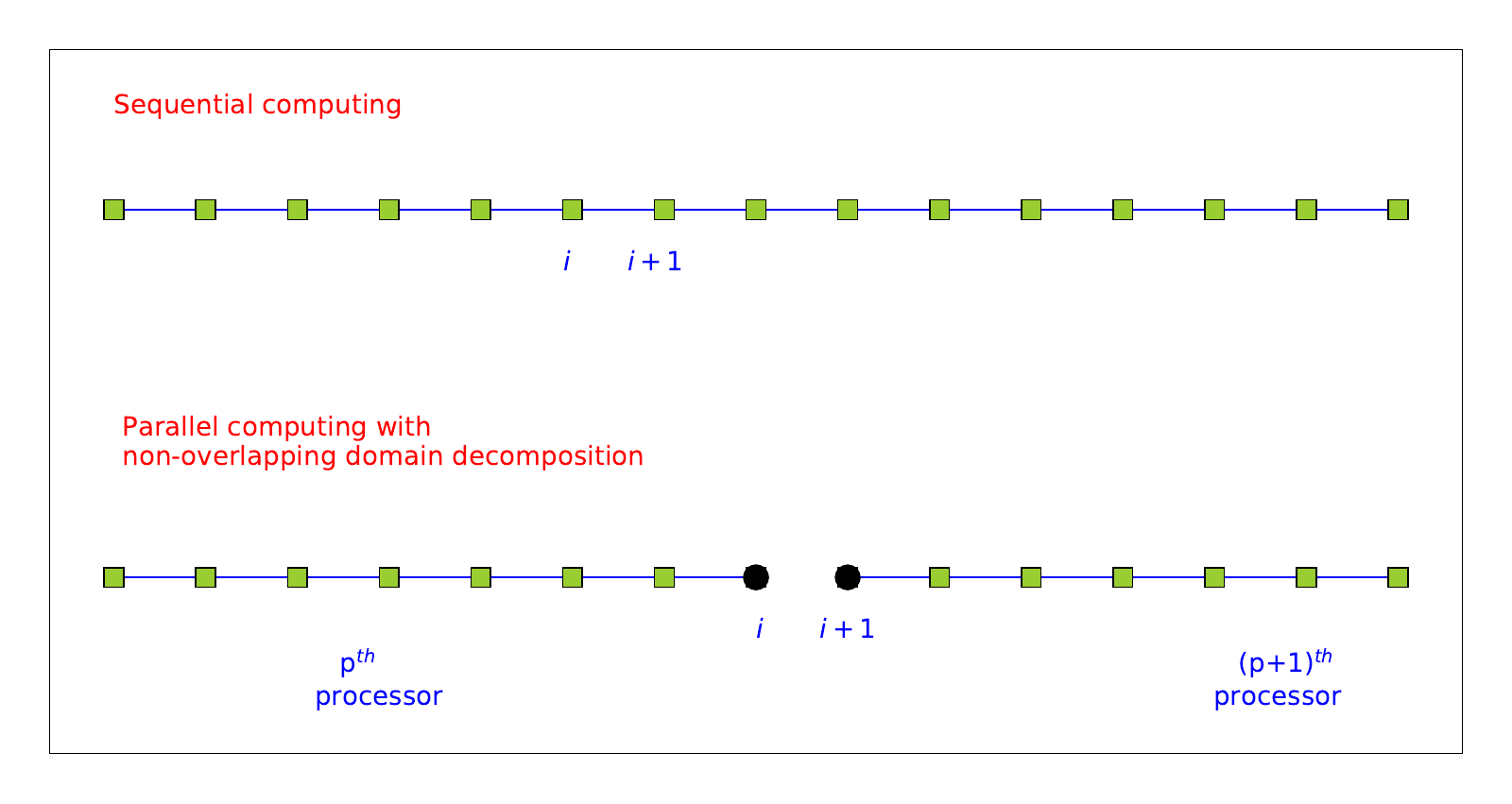}
 \caption{A 1D schematic of non-overlapping domain decomposition of computational domain used in the present computing.}
 \label{Fig:NOHAPSchematic}
 \end{figure*}

Finite difference compact schemes exhibit superior spectral resolution \cite{BhumkarSheuSeng14, B125_28} than the explicit schemes of same order, which motivated researchers to use them in many DNS and large eddy simulations \cite{SenguptaBhau11,B125_20,B125,SumanSivaTekrBhau19,FangGaoMoulEmer19,Kim07,KimSand12,VisbalRizz02,KoutsavdisBlaiLyri00}. The approximation of $m^{th}$ derivative of a function $f$ at $i^{th}$ point using a compact scheme is represented as,
\begin{equation}
 \sum^{n_{l}}_{k=-n_{l}} r_{k,i} f^{(m)}_{i+k} = \sum^{n_{r}}_{k=-n_{r}} s_{k,i} f_{i+k}.
 \label{Eq:CS:GenralForm}
\end{equation}
The coefficients of compact schemes ($r$, $s$) derived for uniform grid spacing are independent of spatial location ($i$). The compact schemes can not be used at the first $max(n_l,n_r)$ points from the boundaries of computational domain and require boundary closure schemes as designed in \cite{B125_28,CarpenterNordGott98}. The compact schemes form a coupled system of equations, which can be represented as,
\begin{equation}
 \left[ A \right] \{ f' \} = \left[ B \right] \{ f \}. \label{Eq:CS:AfpBf}
\end{equation}
In many compact schemes \cite{BhumkarSheuSeng14, B125_28,SharmaSengRajpSamu17}, the $\left[A\right]$ and $\left[B\right]$ matrices are tri- and penta-diagonal matrices, respectively. The non-zero entries of $\left[A \right]$ and $\left[B \right]$ of OUCS3 scheme with $\eta = 0$ \cite{HarasTaas94, B125_28} for the $i^{th}$ node is given as,
 \begin{equation*}
 A_{i,i\pm 1} = 0.3793894912, ~~~~~ A_{i,i} = 1
 \end{equation*}
\begin{equation*}
 B_{i,i\pm 2} = 0.045801298/h, ~~~~~ B_{i,i\pm 1} = 0.7877868952/h, ~~~~~ B_{i,i} = 0
\end{equation*}
The derivatives at a point depend on the derivatives at the neighboring points, which are obtained from the solution of Eq.~\eqref{Eq:CS:AfpBf}. The tri-diagonal matrix algorithm \cite{Thomas49} is used to solve the system of equations and get the derivatives at the grid points. The one-dimensional schematic of a non-overlapping domain decomposition used here is shown in Fig.~\ref{Fig:NOHAPSchematic}. The $i^{th}$ and $(i+1)^{th}$ grid points represent interior grid points in the computational domain, which are now the last and the first grid points of the $p^{th}$ and $(p+1)^{th}$ processors, respectively in the parallel computing. The system of equations given in Eq.~\eqref{Eq:CS:AfpBf} have to be decoupled at $i^{th}$ and $(i+1)^{th}$ points so that, $p^{th}$ and $(p+1)^{th}$ processors can work in parallel, and solve the respective system of equations. An explicit scheme is to be used at the sub-domain boundaries in each processor to decouple the system of equations, which is the proposed sub-domain closure. The difference in the spectral resolution of the compact scheme used in the interior of the domain and the explicit scheme used at the sub-domain boundaries alters the numerical properties at a few grid points near the sub-domain boundaries \cite{B124}. This is the source of additional error in parallel computing of derivatives using compact schemes, a topic which has not been addressed except in \cite{B124}.

There have been efforts in the past to reduce this error due to the parallelization. In \cite{B39}, the domain is discretized with overlap points at the sub-domain boundaries, and one-sided implicit closure is used at the sub-domain boundaries. A large number of overlap points reduce the parallelization error, but which also affects the parallel performance of the simulation. There are also efforts in developing parallel compact schemes without requiring large overlap points. Such methods require one or two overlapping points. The use of an explicit central scheme at the extended sub-domain boundaries reduces bias at the sub-domain boundary point. However, instead of increased accuracy of the closure schemes, discontinuous resolution at the sub-domain boundaries has been noted as a source of error \cite{B124}. To improve the efficiency of large-scale simulations, one would like to decompose the domain with fewer overlap points without compromising the accuracy by filtering the solution \cite{KoutsavdisBlaiLyri00, VisbalRizz02, B39}. Kim \cite{Kim07} used an optimized boundary closure scheme, which was extended in \cite{KimSand12} by including {\it halo nodes} near sub-domain boundaries. Fang {\it et al.} \cite{FangGaoMoulEmer19}, have proposed a parallel compact scheme without overlap points using a similar approach in \cite{KellerKlok13}, with an explicit scheme at the sub-domain boundary of the same order. However, these methods can not fully eliminate errors due to the parallelization of compact schemes. Moreover, the sub-domain closures present in \cite{KellerKlok13, KimSand12, FangGaoMoulEmer19} are specifically designed to work with a chosen compact scheme. In contrast, the NOHAP sub-domain closure \cite{B124} used in the present work does not require overlapping points at the sub-domain boundaries. Computationally this is equivalent to solving the sub-domain boundary point as equivalent to an interior point. Thus, this is equivalent to treating the full domain as part of a single domain as used in sequential computing, retaining the ability to control the error decided upon by machine precision. This is explained next.

For any compact scheme for the first derivative, the equivalent explicit scheme can be obtained as \cite{B124},
\begin{equation}
 \{f'\} = \left[ C \right] \{f\} ~~~~~~~~~~~\text{ where, } \left[C\right] = \left[A\right]^{-1}\left[B\right], \label{Eq:CS:fpCf}
\end{equation}
\noindent and derivatives at $i^{th}$ point is obtained as,
\begin{equation}
 f'_i = \sum_{k=-n_b}^{n_b} C_{ik} f_{i+k}. \label{Eq:CS:fpCf_i}
\end{equation}

The coefficients of the equivalent explicit OUCS3 scheme at an interior node are given in the Appendix. The magnitude of coefficients of $\left[C\right]$ matrix decay exponentially \cite{B124} with respect to diagonal. The semi-bandwidth $(n_b)$ of a compact scheme for double-precision computations is the nodal location from the diagonal of $\left[C\right]$ matrix beyond which the coefficients are less than $10^{-16}$. The $n_b$ of OUCS3 scheme is found to be $48$~\cite{B124}. The solution of Eqs.~\eqref{Eq:CS:AfpBf} and ~\eqref{Eq:CS:fpCf_i} with correct $n_b$ are identical up to machine precision level at which differences arise due to round-off error in computing. The solution via equivalent explicit scheme given in Eq.~\eqref{Eq:CS:fpCf_i} requires $96$ floating-point operations. This is used only to obtain derivatives at the sub-domain boundaries and decouple the system of equations in Eq.~\eqref{Eq:CS:AfpBf}. Then processors independently solve Eq.~\eqref{Eq:CS:AfpBf} to obtain derivatives from second to second-last points in each sub-domain. The NOHAP sub-domain closure is obtained from the compact scheme used in the interior domain. Thus, the spectral resolution is unchanged across the sub-domain boundaries. The NOHAP sub-domain closure eliminates the error due to parallelization completely up to machine precision level \cite{B124}. This methodology of sub-domain boundary closure can be used with any parallel schemes to remove the error at sub-domain boundaries for the chosen precision level. This is established with the help of the DNS of RTI next using the OUCS3 scheme.

\begin{figure*}
\centering
\includegraphics[width=0.8\textwidth]{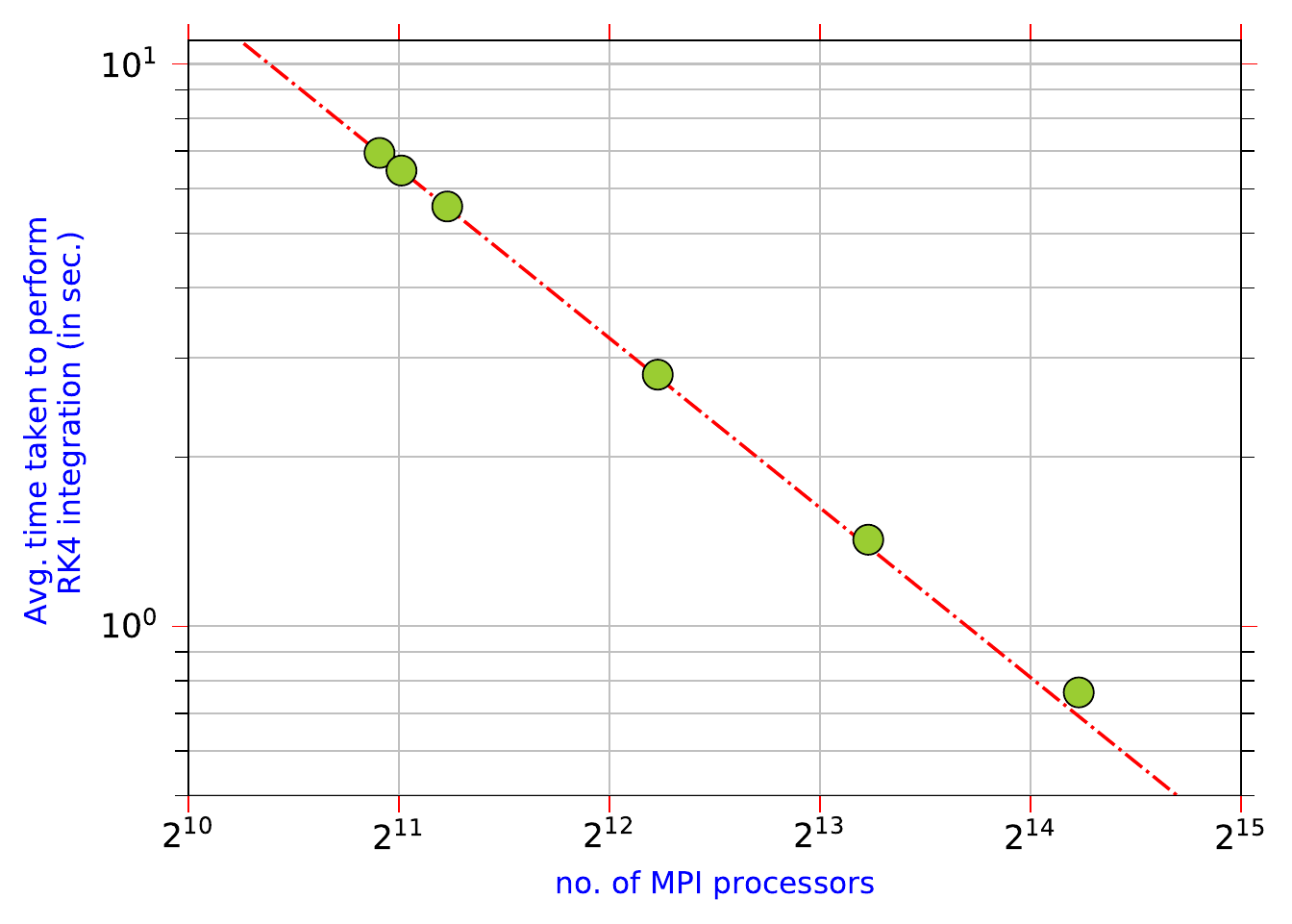}
\caption{Parallel performance of RTI simulations with NOHAP sub-domain closure. Dash-dotted line shows the linear scalability based on the results obtained with the first two sample numbers of processors.}
\label{Fig:Scalability}
\end{figure*}
The parallel performance of NOHAP closure is evaluated from the 3D RTI simulations using $1920$, $2064$, $2400$, $4800$, $9600$, and $19200$ processors. All the simulations are performed in the Cray XC40 machine with Intel Haswell $2.5~GHz$ CPUs that are installed at \href{http://www.serc.iisc.ac.in}{SERC, IISc, Bangalore, India}. The computing node contains $24$ CPUs and $128~GB$ RAM. Floating-point computations are performed with double-precision accuracy. The simulations require $\approx 3.5~TB$ of RAM to solve the 3D compressible NSE in the chosen grid size with ($1280 \times 2560 \times 1280 \approx 4.19$ billion points), and the grid size is kept the same for the tests. The code does not perform writing/reading data to/from the file system during the test.
The time taken to perform time integration is shown as a function of the number of processors in Fig.~\ref{Fig:Scalability}. The dash-dotted line in the figure shows the linear scalability based on the results obtained with $1920$ and $2064$ processors. The 3D RTI simulations with NOHAP closure display linear scaling up to $19200$ processors. This is more than $99 \%$ of parallel efficiency with respect to $1920$ processors result observed up to $4800$ processors, and $91.08 \%$ of performance is retained while using $19200$ processors.
The results demonstrate an excellent parallel efficiency and linear scaling of the solver with NOHAP closure. A careful design and implementation of NOHAP closure allow one to retain the high accuracy of compact schemes along with a good parallel performance, demonstrating the potential of NOHAP closure for even exascale computing.

\begin{figure*}
\centering
\includegraphics[width=0.9\textwidth]{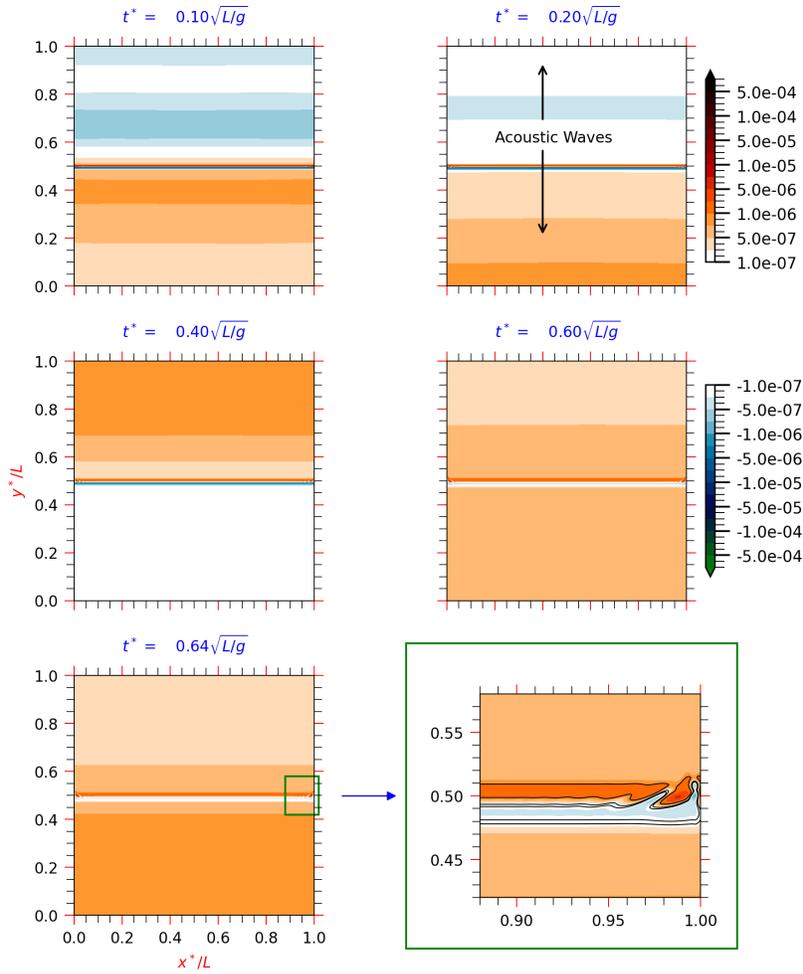}
\caption{Initial evolution of disturbance pressure in $z = 0.5$ plane at indicated times instants. Zoomed view in the bottom-right frame shows the visible vortical structure of evolving RTI.}
\label{Fig:Prsr2D}
\end{figure*}

\begin{figure*}
\centering
\includegraphics[width=0.9\textwidth]{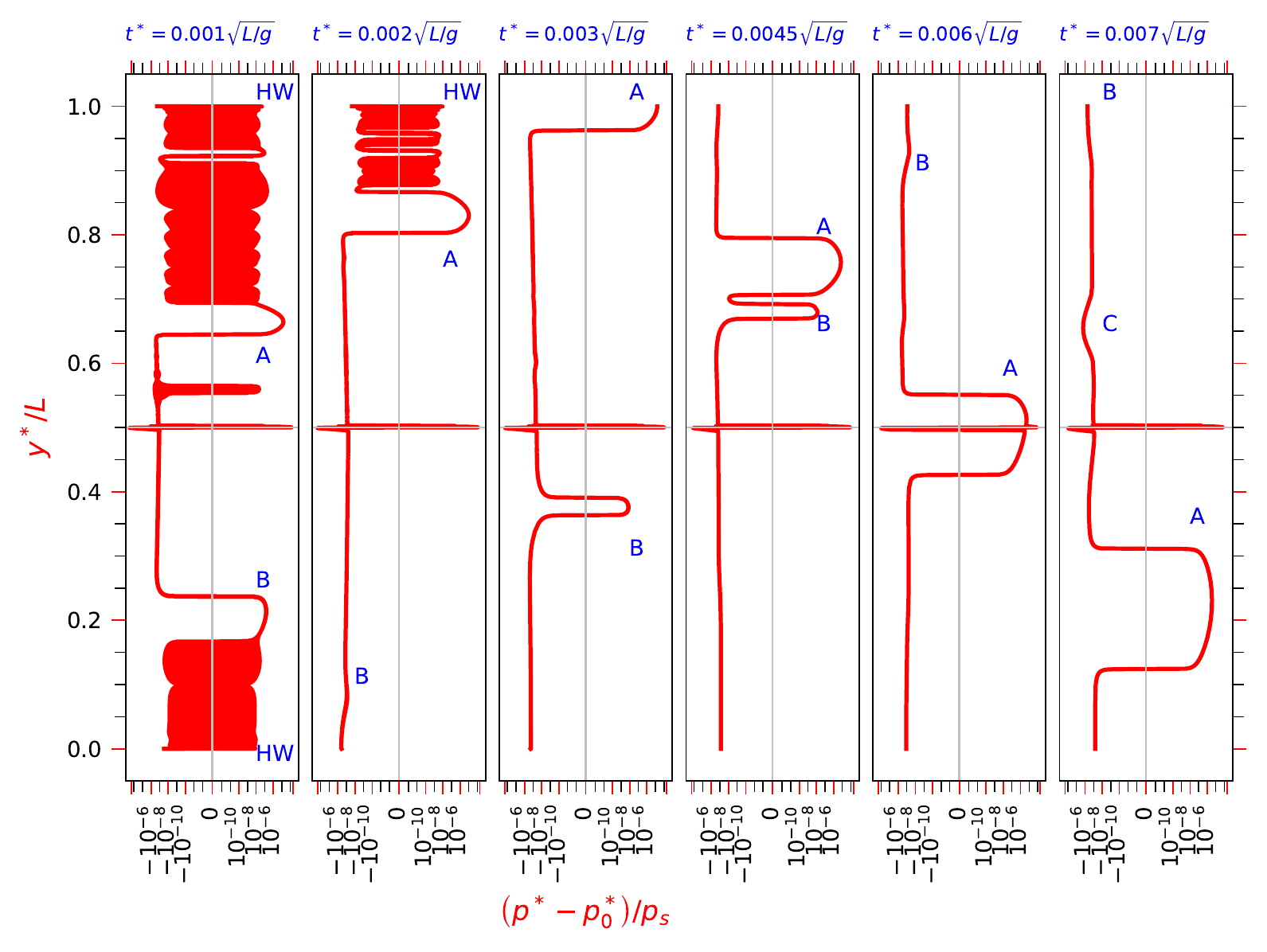}
\caption{Initial evolution of disturbance pressure in $x^*=z^*=0.5L$ location at indicated times instants is plotted as a function height. The high wavenumber oscillations at initial time instants are denoted by $HW$.}
\label{Fig:Prsr1D}
\end{figure*}

\begin{figure*}
\centering
\includegraphics[width=0.9\textwidth]{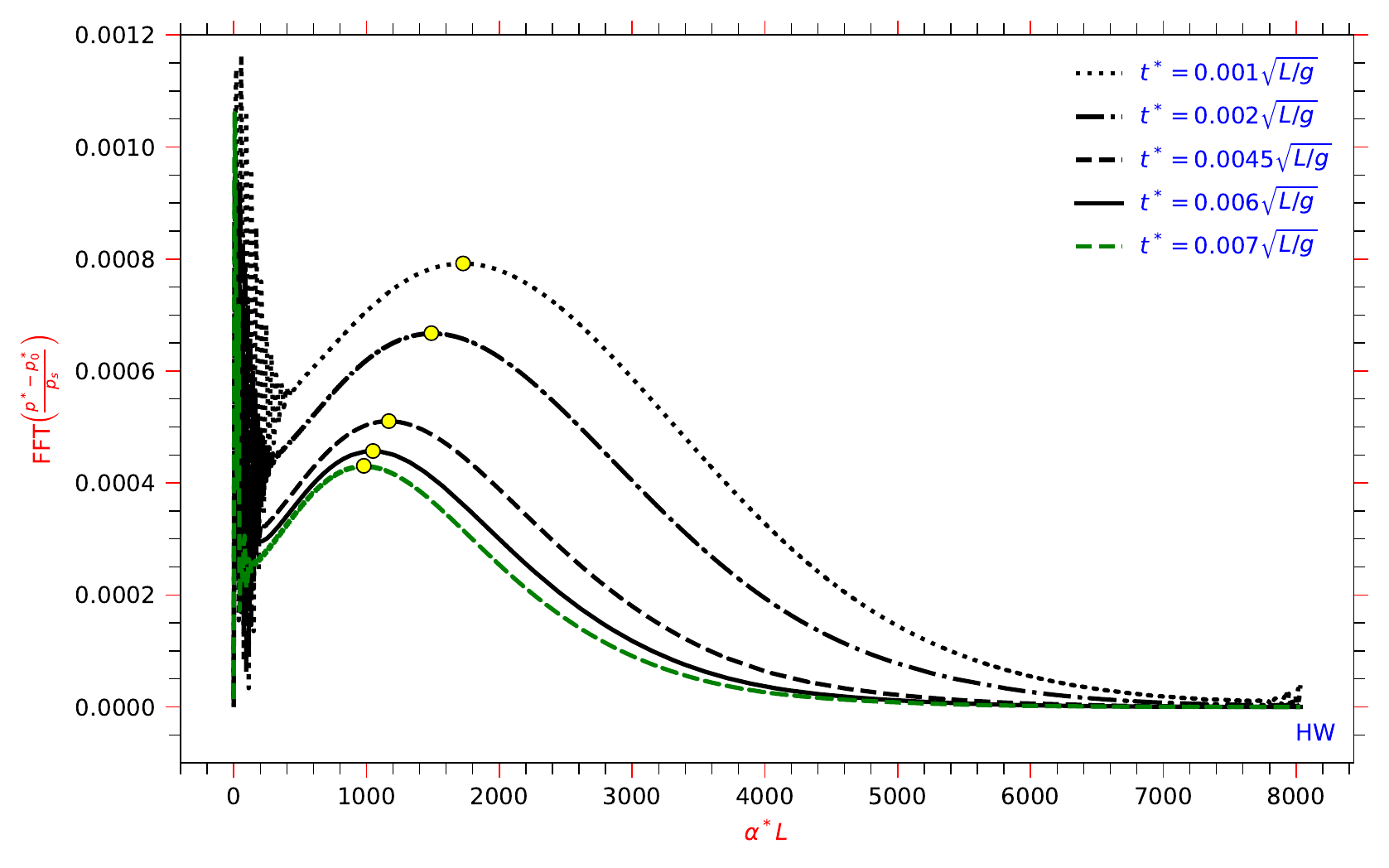}
\caption{Fourier transform of disturbance pressure in Fig~\ref{Fig:Prsr1D} for $x^*=z^*=0.5L$ location at indicated times instants. The circles denote the central wavenumber and peak amplitude of the pressure waves.}
\label{Fig:Prsr1D_FFT}
\end{figure*}

\begin{figure*}
\centering
\includegraphics[width=0.9\textwidth]{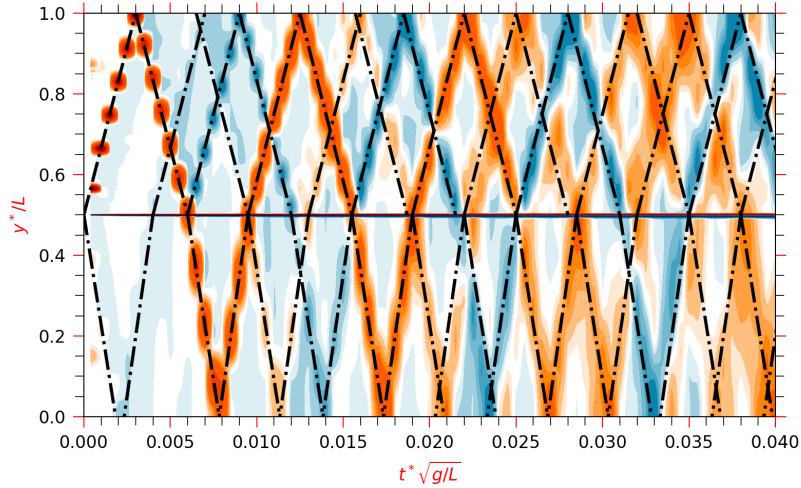}
\caption{The contours of disturbance pressure in $x^*=z^*=0.5L$ location are plotted in $(y,t)-$plane at indicated times instants.}
\label{Fig:Prsr1D_yt}
\end{figure*}

\begin{figure*}
   \centering
   \includegraphics[width=0.9\textwidth]{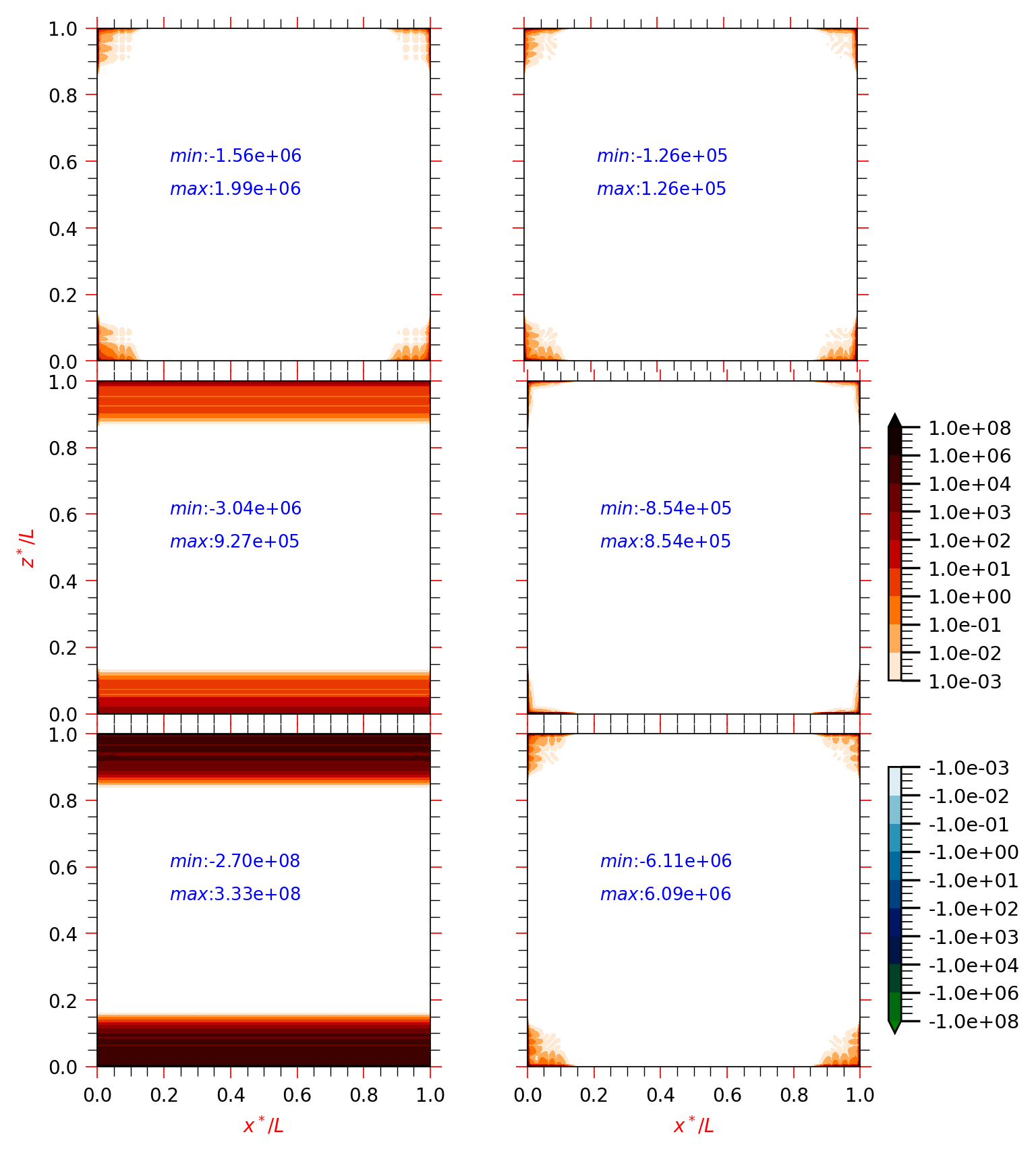}
   \caption{The first, second and the third terms on the right-hand side of Eq.~\eqref{eqn12_1} at $y^* = 0.5L$ and $t^* = 6.25\times10^{-5}$ are plotted in the top, middle and bottom rows of the figure, respectively. The first and second columns show the $x-$ and $y-$ components of each term.}
   \label{Fig:DwDt_RHS}
   \end{figure*}

\section{Acoustic Trigger for the RTI}
The present results are computed using ($20\times48\times20$) = 19,200 processors and NSE solver takes $\approx 0.85 s$ per time step. Onset of RTI is captured by solving NSE up to $t^* = 0.7 \sqrt{L/g}$, which requires $3,686,400$ core-hours of computation with Intel Haswell $2.5~GHz$ CPUs installed at \href{http://www.serc.iisc.ac.in}{SERC, IISc, Bangalore, India}. The OUCS3 scheme \cite{HarasTaas94, B125_28} with NOHAP sub-domain closure is used to discretize the convective fluxes, and the second-order explicit finite difference scheme is used to discretize the viscous flux and other gradients. The four-stage, fourth-order Runge-Kutta scheme \cite{B125_28} is used to perform time integration. The implicit 3D filter \cite{BhumkarSeng11, B125_28} with 0.135 as filtering coefficient is applied at every $50$ timesteps is applied to eliminate high wavenumber oscillations in the simulation, and the results are discussed in this section.

At the initial unperturbed state, one notes that at the interface of the compartments, the density and pressure gradient act in the upward and downward directions, respectively. The impulsive removal of partition between the air masses propagates the information as pressure waves in the direction normal to the interface as shown in Fig.~\ref{Fig:Prsr2D}. The nondimensionalized instantaneous pressure field is subtracted from the initial hydrostatic pressure and plotted for $(z^*=0.5L)$-plane at indicated times in the figure. The pressure waves are reflected at the top and bottom walls, which travel towards the interface. The interface further partially reflects the traveling pressure waves, and a complex wave system is formed. The magnitude of the disturbance pressure is in the order of $10^{-6}$, which is many orders of magnitude smaller than the hydrostatic pressure. The error triggered by an improper parallelization of compact schemes is shown to be of a similar order of magnitude \cite{B124}. In contrast, the NOHAP closure removes parallelization errors at the sub-domain boundaries and accurately computes this multi-physics problem. The zoomed view in the bottom-right frame of the figure shows a visible vortical structure of the evolving RTI at the indicated time.

Figure~\ref{Fig:Prsr2D} shows that the disturbance pressure field at the initial times is varying only in the direction normal to the interface. Thus, the variation of disturbance pressure at $x^*=z^*=0.5L$ is plotted in Fig.~\ref{Fig:Prsr1D} at the indicated time instants to understand the evolution of disturbances. The discontinuity at the interface excites wavenumbers with an amplitude proportional to $1/\alpha$ \cite{IFTT}. The pressure waves' genesis is from the interface, which travel towards the top and bottom walls. The high wavenumber components marked as ($HW$) in the frames diffuse quickly over time. The components identified as $A$ and $B$ in the figure propagate towards the top and bottom walls, respectively, and are reflected by the walls. The component $A$ is partially reflected by the interface at $t = 0.0065$, and component $C$ is created, which propagates in the opposite direction, as shown in the last frame of the figure.

The corresponding Fourier transform of disturbance pressure field is shown in Fig.~\ref{Fig:Prsr1D_FFT} for the indicated time instances. The peaks at low wavenumbers ($\alpha^* < 600/L$) in Fig~\ref{Fig:Prsr1D_FFT} correspond to the variation of disturbance pressure at the extreme $y$ locations \cite{IFTT} in Fig~\ref{Fig:Prsr1D}. The high wavenumber components ($\alpha^* > 7600/L$) created by the discontinuity in the initial condition diffuse over time. The central wavenumber and corresponding magnitude of the pressure wave are marked with circles in Fig.~\ref{Fig:Prsr1D_FFT}. The strength and wavenumber of the pressure wave decay over time.

The disturbance pressure at $x^*=z^*=0.5L$ is plotted in the $(y,t)$-plane, in Fig~\ref{Fig:Prsr1D_yt}. The peaks of the pressure waves at early times are connected with dash-dotted lines in the figure. The slope of the dash-dotted line provides the speed of propagation of the pressure waves in the fluid domain. The slope of the lines are equal to the local speed of sound, which is given as, $$\frac{a^*}{U_s} = \frac{\sqrt{T}}{M} = 200.46 m/s ~~\text{and}~~347.21 m/s$$ on the top and bottom side of the interface, respectively. The dispersion relation of the linear convection equation ($\omega_0 = a \alpha$) \cite{B125_28} provides the range of frequency of acoustic waves as, $0.8~MHz$ - $3.73~MHz$. These ultrasonic sound waves serve as a deterministic mechanism that triggers the RTI of air masses separated by an unperturbed interface and ensures the reproducibility of the simulations.

In the 2D RTI simulation \cite{B125_21}, it has been noted that the initial 1D pressure pulses become 2D complex wave systems, and one observes the visual signature of vortical structures that leads to the evolution of mixing at the interface. In the present 3D RTI simulation, it is important to ascertain how truly the onset of RTI takes place and its relation to initial ultrasounds reported so far in the above. It has been noted by Chandrasekhar \cite{B125_6} and in visualization experiments \cite{B125_1, B125_7} that the visible onset of RTI is due to vorticity evolution at the interface, and this has been attributed to the baroclinic instability. Thus, in the following section, one is interested in pinpointing the onset of RTI with the evolution of vorticity at the interface plane and whether the other terms of vorticity transport equation also contribute to it.

\begin{center}
   \href{https://youtu.be/hCa_MjOo-j4}{\includegraphics[width=\textwidth]{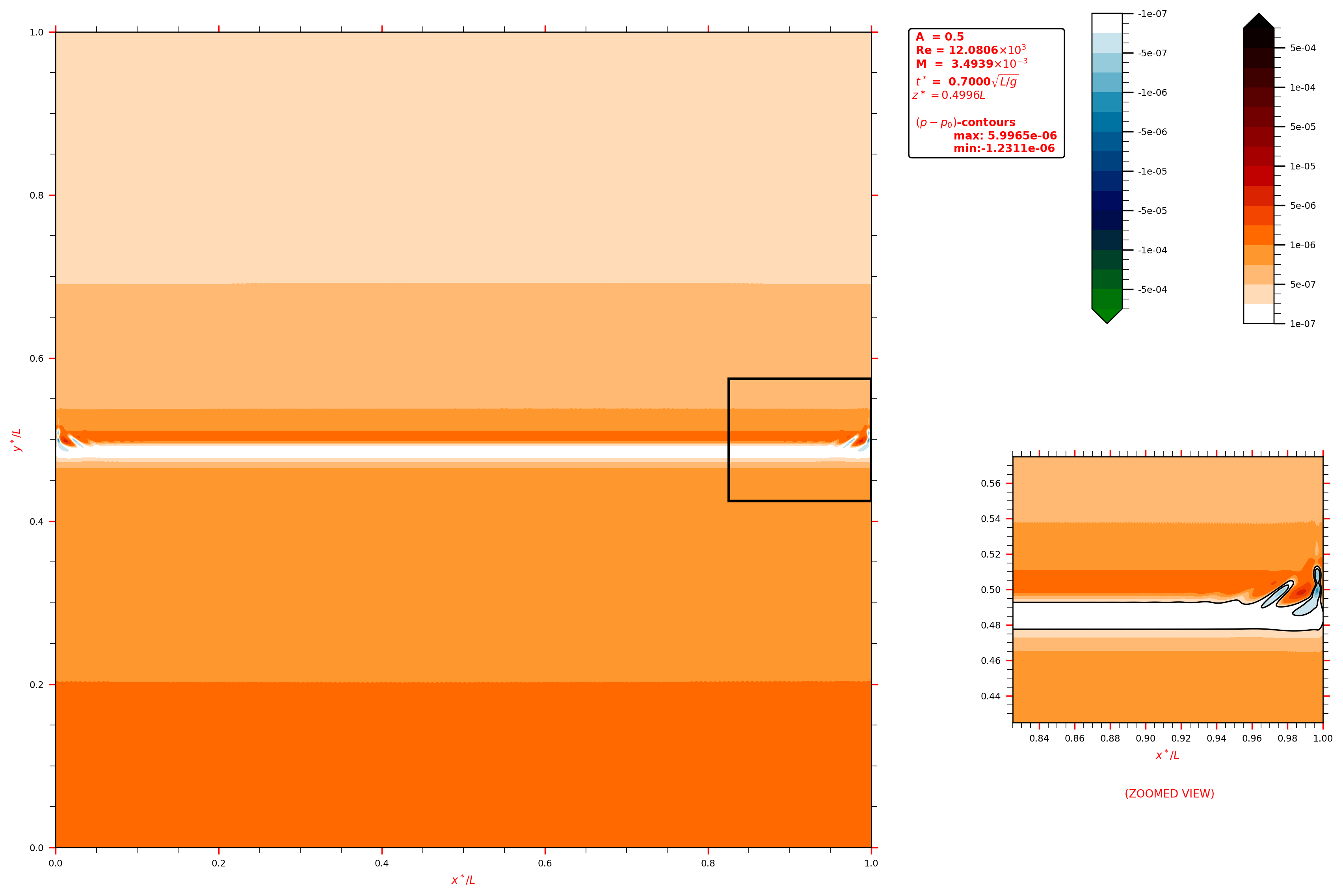}
   \\Video-1 (click caption/figure to play the video): The evolution of disturbance pressure at $z = 0.5$ plane.}
   \label{Video-1}
   \end{center}

\section{Temporal Growth of Disturbances in the Interface Plane}
The evolution of temporal growth of vorticity in the interface plane is investigated using Eq.~\eqref{eqn12_1}. This equation provides the clue of the temporal growth rate of vorticity that is typical of the present simulation based on the specific formulation. The various terms present on the right-hand side of Eq.~\eqref{eqn12_1} contribute to the growth rate of vorticity at the ($y^* = 0.5L$)-plane, which are due to (i) vortex-stretching (typical of 3D formulation), (ii) contribution due to dilation and compression locally by the divergence of the velocity term (due to bulk viscosity contribution) and also (iii) due to misalignment of density gradient from the pressure gradient (baroclinic term). These are present, apart from viscous term contribution an increase of enstrophy \cite{SenguptaSumaSengBhau18}. The above three terms on the right-hand side of Eq.~\eqref{eqn12_1} are plotted for an early time ($100 \Delta t$) in Fig.~\ref{Fig:DwDt_RHS}. These stretching, divergence and baroclinic terms are plotted on the top, middle, and bottom frames, respectively. While the left column shows the $x$-component (which is identical to the $z$-component), the right column shows the $y$ component of each term. The results show that the growth of RTI begins from the corner and edges of the interface, along with the temporal growth of vorticity triggered initially due to the acoustic pulses. The maximum amplitude of each term is shown in each frame, which clearly indicates that the baroclinic term contributes more than $97\%$ to the total growth of vorticity at the interface. The contribution by the bulk viscosity and vortex stretching terms contribute to the rest by almost the same order of magnitude. Thus, the acoustic pulses and their reflections (whose origin can be traced back to the ultrasound created at the onset of the experiment) from the walls cause the misalignment of the pressure and density gradient vectors, leading to baroclinic vorticity production at the interface. The convection created by the generated vorticity enhances the mixing of the heavy and lighter fluids as investigated in \cite{B125_7, B125_21, B125}.
%
The temporal evolution of disturbance pressure at $(z^* = 0.5 L)$-plane is shown in \href{https://youtu.be/hCa_MjOo-j4}{Video-1}. The ultrasonic acoustic waves predominately travel in the direction orthogonal to the initial interface and suffer multiple reflections from the walls and the interface. The frequency and wavenumber of the acoustic waves decay over time. The bottom-right frame in the animation shows the onset of visible structures of RTI due to background acoustic disturbance.

\section{Summary and Conclusions}
The onset mechanisms involved in the buoyancy-driven Rayleigh-Taylor instability (RTI) are numerically investigated in the present work by performing 3D DNS of a flow field consisting of air at two different temperatures (and hence densities), initially separated by a non-conducting partition. The onset of the numerical experiment involves the removal of this partition at $t = 0$. The computations involved solving the 3D NSE for compressible flows by adopting a parallel algorithm which allows one to operate with computing error akin to that of an equivalent sequential computation. This NOHAP strategy for parallelization allowed one to capture very high-frequency acoustic signals in the RTI, which are the harbingers of the onset mechanism, by computing using $\approx$ 4.2 billion mesh points and a time-step of $7.68 \times 10^{-8}$.

The parallel performance of the NOHAP scheme has been demonstrated by a series of 3D RTI simulations using 1920, 2064, 2400, 4800, 9600, and 19,200 processors. A $91.08 \%$ retention of parallel performance has been demonstrated using 19,200 processors, suggesting a linear scaling of the present solver. Pressure waves reflected back and forth from the interface to the top/bottom walls were captured using the error-free parallelization tool of NOHAP. These pressure waves, in turn, form a complex wave system at the interface, which aids the formation of baroclinic vorticity. This vorticity is the seed for the onset of the RTI, as has been reported in prior works \cite{B125_6, TT_CUP, B125}. The high wavenumber components of these acoustic waves diffused quickly with time, and the typical range of frequencies associated with these are between 0.8 to 3.73 MHz. These ultrasonic waves serve as a deterministic trigger for the generation of baroclinic vorticity, and hence the onset of the RTI.

The onset and propagation of RTI are from the junction of the corner and edges of the 3D box simulated with the interface. This is followed by a spatio-temporal growth of the vorticity triggered at these junction points due to the acoustic waves. The further evolution of the instability from this onset to subsequent development of the mixing layer is aided by the convection of the generated vorticity. An analysis of the terms of the vorticity transport equation revealed that baroclinic term contributes to more than $97\%$ of vorticity generation during onset, confirming the commonly adopted viewpoint regarding RTI inception \cite{B125_6, B125_7}.

\section*{Acknowledgments}
The authors are grateful to National Supercomputing Mission, Government of India, Grant No. DST/NSM/R$\&$D{\_}Exascale/2021/17 for sponsoring the project and providing full financial support for the problem solved here as part of an NSM initiative. The authors thank SERC, IISc, Bangalore, India, for providing the computing facility for the simulations.

\section*{Appendix: Coefficients of equivalent explicit OUCS3 scheme at an interior node}
\begin{center}
\begin{tabular}{l|l}
$C_{i,i}$~~ = ~7.4593109467002705e-17/h & \\
$C_{i,i-1}$ = -9.2862503092980486e-01/h & $C_{i,i+1}$ = ~9.2862503092980497e-01/h \\
$C_{i,i-2}$ = ~3.7122308075623583e-01/h & $C_{i,i+2}$ = -3.7122308075623595e-01/h \\
$C_{i,i-3}$ = -1.7057351944964194e-01/h & $C_{i,i+3}$ = ~1.7057351944964191e-01/h \\
$C_{i,i-4}$ = ~7.8376930330317604e-02/h & $C_{i,i+4}$ = -7.8376930330317590e-02/h \\
$C_{i,i-5}$ = -3.6013463448627651e-02/h & $C_{i,i+5}$ = ~3.6013463448627665e-02/h \\
$C_{i,i-6}$ = ~1.6547848252025113e-02/h & $C_{i,i+6}$ = -1.6547848252025120e-02/h \\
$C_{i,i-7}$ = -7.6035808708780402e-03/h & $C_{i,i+7}$ = ~7.6035808708780411e-03/h \\
$C_{i,i-8}$ = ~3.4937740049016432e-03/h & $C_{i,i+8}$ = -3.4937740049016445e-03/h \\
$C_{i,i-9}$ = -1.6053563452027701e-03/h & $C_{i,i+9}$ = ~1.6053563452027705e-03/h \\
$C_{i,i-10}$ = ~7.3764616471103074e-04/h & $C_{i,i+10}$ = -7.3764616471103117e-04/h \\
$C_{i,i-11}$ = -3.3894148544581613e-04/h & $C_{i,i+11}$ = ~3.3894148544581624e-04/h \\
$C_{i,i-12}$ = ~1.5574042956113061e-04/h & $C_{i,i+12}$ = -1.5574042956113064e-04/h \\
$C_{i,i-13}$ = -7.1561264824169580e-05/h & $C_{i,i+13}$ = ~7.1561264824169634e-05/h \\
$C_{i,i-14}$ = ~3.2881729154502238e-05/h & $C_{i,i+14}$ = -3.2881729154502265e-05/h \\
$C_{i,i-15}$ = -1.5108845754007249e-05/h & $C_{i,i+15}$ = ~1.5108845754007259e-05/h \\
$C_{i,i-16}$ = ~6.9423727367186414e-06/h & $C_{i,i+16}$ = -6.9423727367186465e-06/h \\
$C_{i,i-17}$ = -3.1899550766643657e-06/h & $C_{i,i+17}$ = ~3.1899550766643686e-06/h \\
$C_{i,i-18}$ = ~1.4657544008428492e-06/h & $C_{i,i+18}$ = -1.4657544008428501e-06/h \\
$C_{i,i-19}$ = -6.7350038228022020e-07/h & $C_{i,i+19}$ = ~6.7350038228022094e-07/h \\
$C_{i,i-20}$ = ~3.0946710081222926e-07/h & $C_{i,i+20}$ = -3.0946710081222952e-07/h \\
$C_{i,i-21}$ = -1.4219722661609411e-07/h & $C_{i,i+21}$ = ~1.4219722661609419e-07/h \\
$C_{i,i-22}$ = ~6.5338290255213388e-08/h & $C_{i,i+22}$ = -6.5338290255213468e-08/h \\
$C_{i,i-23}$ = -3.0022330780052853e-08/h & $C_{i,i+23}$ = ~3.0022330780052886e-08/h \\
$C_{i,i-24}$ = ~1.3794979053572500e-08/h & $C_{i,i+24}$ = -1.3794979053572518e-08/h \\
$C_{i,i-25}$ = -6.3386633263977726e-09/h & $C_{i,i+25}$ = ~6.3386633263977817e-09/h \\
$C_{i,i-26}$ = ~2.9125562720600853e-09/h & $C_{i,i+26}$ = -2.9125562720600894e-09/h \\
$C_{i,i-27}$ = -1.3382922551807741e-09/h & $C_{i,i+27}$ = ~1.3382922551807757e-09/h \\
$C_{i,i-28}$ = ~6.1493272334615784e-10/h & $C_{i,i+28}$ = -6.1493272334615877e-10/h \\
$C_{i,i-29}$ = -2.8255581154121207e-10/h & $C_{i,i+29}$ = ~2.8255581154121233e-10/h \\
$C_{i,i-30}$ = ~1.2983174192011675e-10/h & $C_{i,i+30}$ = -1.2983174192011691e-10/h \\
$C_{i,i-31}$ = -5.9656466161741746e-11/h & $C_{i,i+31}$ = ~5.9656466161741849e-11/h \\
$C_{i,i-32}$ = ~2.7411585967141740e-11/h & $C_{i,i+32}$ = -2.7411585967141791e-11/h \\
$C_{i,i-33}$ = -1.2595366329557733e-11/h & $C_{i,i+33}$ = ~1.2595366329557751e-11/h \\
$C_{i,i-34}$ = ~5.7874525452822067e-12/h & $C_{i,i+34}$ = -5.7874525452822164e-12/h \\
$C_{i,i-35}$ = -2.6592800945607446e-12/h & $C_{i,i+35}$ = ~2.6592800945607491e-12/h \\
$C_{i,i-36}$ = ~1.2219142301377044e-12/h & $C_{i,i+36}$ = -1.2219142301377066e-12/h \\
$C_{i,i-37}$ = -5.6145811374549513e-13/h & $C_{i,i+37}$ = ~5.6145811374549634e-13/h \\
$C_{i,i-38}$ = ~2.5798473061003951e-13/h & $C_{i,i+38}$ = -2.5798473061004002e-13/h \\
$C_{i,i-39}$ = -1.1854156097938961e-13/h & $C_{i,i+39}$ = ~1.1854156097938984e-13/h \\
$C_{i,i-40}$ = ~5.4468734045624554e-14/h & $C_{i,i+40}$ = -5.4468734045624668e-14/h \\
$C_{i,i-41}$ = -2.5027871777804704e-14/h & $C_{i,i+41}$ = ~2.5027871777804755e-14/h \\
$C_{i,i-42}$ = ~1.1500072045029490e-14/h & $C_{i,i+42}$ = -1.1500072045029515e-14/h \\
$C_{i,i-43}$ = -5.2841751074557042e-15/h & $C_{i,i+43}$ = ~5.2841751074557184e-15/h \\
$C_{i,i-44}$ = ~2.4280288381604572e-15/h & $C_{i,i+44}$ = -2.4280288381604631e-15/h \\
$C_{i,i-45}$ = -1.1156564495035025e-15/h & $C_{i,i+45}$ = ~1.1156564495035057e-15/h \\
$C_{i,i-46}$ = ~5.1263366141144072e-16/h & $C_{i,i+46}$ = -5.1263366141144200e-16/h \\
$C_{i,i-47}$ = -2.3555035327322291e-16/h & $C_{i,i+47}$ = ~2.3555035327322350e-16/h \\
$C_{i,i-48}$ = ~1.0823317527447454e-16/h & $C_{i,i+48}$ = -1.0823317527447481e-16/h
\end{tabular}
\end{center}
\bibliographystyle{elsarticle-num}
\bibliography{RTI_Acoustics}
\end{document}